\documentclass[showpacs,preprint]{revtex4}
\usepackage{amssymb}
\usepackage{graphicx}
\usepackage{dcolumn}
\usepackage{bm}

\begin{document}

\title{Radiative Losses in Plasma Accelerators}
\author{I.~Kostyukov}
\email{kost@appl.sci-nnov.ru}
\author{E.~Nerush}
\affiliation{Institute of Applied Physics, Russian Academy of Science,
  46 Uljanov St. 603950 Nizhny Novgorod, Russia}
  \author{A.~Pukhov}
\address{  Heinrich-Heine-Universitat Duesseldorf,
40225 Duesseldorf, Germany}
\date{\today}

\begin{abstract}
We investigate the dynamics of a relativistic electron in a strongly nonlinear plasma wave in terms of classical mechanics by taking into account the action of the radiative reaction force. The two limiting cases are considered. In the first case where the energy of the accelerated electrons is low, the electron makes many betatron oscillations during the acceleration. In the second case where the energy of the accelerated electrons is high, the betatron oscillation period is longer than the electron residence time in the accelerating phase. We show that the force of radiative friction can severely limit the rate of electron acceleration in a plasma accelerator.
\end{abstract}

\maketitle

\section*{1. Introduction}

New acceleration methods that provide a high rate of energy gain by particles are currently being developed
to generate charged particles with very high energies. One of the promising methods is the acceleration
of charged particles in a plasma wave excited by a short intense laser pulse or a dense bunch of relativistic electrons \cite{Tadjima}-\cite{Esarey}. The longitudinal electric field in a plasma wave can reach huge strengths. Such a field is capable of accelerating electrons to very high energies. 

Recently, substantial progress has been made in generating short quasi-monoenergetic beams of
ultrarelativistic electrons in laser plasma \cite{Katsoleas}. One of the models \cite{Pukhov2002,Kostyukov2004} that describe the generation of a quasimonoenergetic beam of ultrarelativistic particles suggests that this generation is associated with the transition
to a strongly nonlinear regime of laser pulse- plasma interaction. In this regime, the periodic plasma
wave in the wake of a laser pulse transforms to a plasma cavity without electrons. Plasma electrons, along with external beam electrons, can be trapped by the cavity and accelerated to very high energies.

It should be noted that, apart from a strong accelerating electric field, significant focusing transverse
fields act on the electrons. These forces may be comparable in intensity to the accelerating force. The action of the focusing forces leads to the excitation of betatron electron oscillations across the acceleration direction. As a result of these oscillations, the relativistic electrons intensely emit electromagnetic waves \cite{Kiselev2004}. Just as in conventional accelerators, the losses through radiation can greatly reduce the electron acceleration efficiency in a plasma cavity.

Since the properties of the electromagnetic radiation from an electron are determined mainly by the action of
the focusing forces, when these properties are studied, the action of the accelerating longitudinal force may be disregarded, i.e., the electron dynamics may be considered in an infinitely long ion channel. In this case, the shape of the electron trajectory in the channel is nearly sinusoidal \cite{essarey} and is defined by the betatron frequency.
\begin{equation}
\Omega = \frac{ \omega_p } {\sqrt{2 \gamma}},
\label{betatron-frequency}
\end{equation}
\noindent
where $\omega _p = \left( {4\pi e^2n_0 / m} \right)^{1 / 2}$ is the electron plasma frequency, $\gamma $ is the relativistic gamma factor of the electron, $n_0 $ is the unperturbed plasma density, $e$ is the electron
charge, and $m$ is the electron mass.

The trajectory of an electron in an ion channel is similar to the helical trajectory of an electron in a constant magnetic field. The electron gyration in a magnetic field is defined by the cyclotron frequency
$\omega _c = e H/ (m c)$ \cite{landau}, where $H$ is the magnetic field strength and $c$ is the speed of light. As a result, the electromagnetic radiation spectrum of a relativistic electron \cite{essarey,kostyukov2003} that makes betatron oscillations in an ion channel is similar to the synchrotron radiation spectrum \cite{landau} of an electron in a magnetic field. In the case of a relativistic
transverse electron momentum, the spectrum is quasicontinuous. The synchrotron radiation spectrum is
defined by a critical frequency that is a function of the betatron frequency for the radiation from the ion
channel
\begin{equation}
\hslash \omega _{cr}=\frac{3}{2}\gamma ^{3} \hslash r_0 \Omega ^{2} / c \simeq
5\times 10^{-24}\gamma^{2}n_{0}\left[ cm^{-3}\right] r_{0} \left[ \mu
m\right] keV,  
\label{crytical-frequency}
\end{equation}
\noindent
where $r_0$ is the amplitude of the betatron electron oscillations in the channel. For frequencies much lower than the critical frequency, the radiated energy increases as $\sim \omega ^{2/3}$, reaching its maximum near  $\sim 0.29\omega _{cr}$, and decreases exponentially at  $\omega > \omega _{cr}$.

By analogy with the synchrotron radiation of an electron in a magnetic field, which was studied in detail
in \cite{sokolov}, the effects related to the action of the radiative reaction force on an electron moving in an ion channel can be estimated. When the radiated photon energy becomes comparable to the electron energy (
$mc^2 \gamma < \hslash \omega _{cr}$), the electron dynamics becomes quantum one. The limiting electron energy at which the quantum effects become significant can be estimated using Eq.~(\ref{crytical-frequency}) 
\begin{equation}
\gamma \simeq \frac{4}{3} \frac{m c^4} {\hslash c r_0 \omega_{p}^{2}}.
\label{quantum}
\end{equation}
\noindent
Strictly speaking, the quantum nature of the radiation
can lead to a broadening of the electron orbit even at energies below the limiting value of (\ref{quantum}) \cite{sokolov}. However, these effects, which are significant for the long-term electron dynamics characteristic of conventional accelerators, are insignificant for the acceleration in plasma, because the interaction time is short.

The space-time distribution of the electromagnetic fields in a plasma cavity are discussed in Section.~2. We
derive the equations that describe the dynamics of a relativistic electron in a plasma cavity by taking into
account the radiative reaction force. We obtain an estimate for the limiting energy of the accelerated electron that is limited by radiative losses. In Section~3, the electron dynamics in a plasma cavity is considered by taking into account the radiative reaction force in the case where the electron makes many betatron oscillations while flying through the cavity. We investigate the evolution of the electron energy as a result of radiative losses in an infinitely long ion channel. In Section~4, the electron dynamics is analyzed in the opposite limit of high energies, where the electron without the action of the radiative force makes no more than one betatron oscillation while flying through the cavity. In Section~5, we present the results of our numerical simulations of the acceleration of a relativistic electron bunch in a
plasma cavity by taking into account the radiative reaction force using a hybrid particle-in-cell code
(software package). The results obtained are discussed in Section~6.

\section*{2.  ELECTRON DYNAMICS WITH THE RADIATIVE REACTION FORCE}

As follows from our numerical simulations, the shape of the plasma cavity generated in the wake of a
laser pulse is nearly spherical. The velocity of the plasma cavity,  $v_l $, is equal to the group velocity of the laser pulse and is close to the speed of light. The ions may be considered to be stationary, since the characteristic ion response time  $\tau _i \approx \left( {4\pi  e^2n_0 / m_i } \right)^{-1/2}$ for typical
interaction parameters is much longer than $R / c$, where $m_i $ is the ion mass, $R$ is the cavity radius, 
$c$ is the speed of light. Thus, the electron density in the cavity is zero, while the ion density is
 $n_0 $. In this case, the potential distribution inside the cavity is   \cite{Kostyukov2004}
\begin{equation}
\varphi = - A_x = \frac{\xi ^2 + y^2 + z^2}{8},
\label{potential}
\end{equation}
\noindent
where $\xi = x - v_l t$, $\varphi $ is the scalar potential, ${\rm {\bf A}}$ is the vector potential, and the gauge $\varphi = - A_x $is written. In what follows, we use dimensionless quantities where the
time, the velocity, the length, the electromagnetic field strength, and the electron density, $n$, are normalized to  $1 / \omega _p $, $c / \omega _p $, $mc\omega _p /|e|$, and $n_0 $ , respectively.
The laser pulse and the plasma cavity are assumed to propagate along the $x$ axis. The space-time
distribution of the quasi-static plasma fields inside the cavity is described by a linear function of the coordinates and time  \cite{Kostyukov2004}
\begin{equation}
E_x = \xi / 2,
\quad
E_y = - B_z = y / 4,
\quad
E_z = B_y = z / 4.
\label{fields}
\end{equation}
\noindent
It is important to note that a similar distribution of the electromagnetic fields is also observed in a strongly nonlinear plasma wave excited by an electron beam \cite{barov}-\cite{Mori}.

Let us first consider the electron acceleration in the cavity without the radiative reaction force. For the sake of simplicity, we assume that the electron trajectory lies in the $z = 0$ plane and that the electron is accelerated along the $x$ axis, making betatron oscillations along the $y$ axis, with $p_x \gg p_y \gg 1$, 
where $p_x $ and $p_y $ are the longitudinal and transverse electron momenta, respectively. In this case, it is easy to find the trajectory of the electron acelerated in the cavity \cite{Kostyukov2004} 
\begin{equation}
\xi \approx - R + \frac{t}{2\gamma _l^2 },
\quad
y \approx r_0 \left( {\frac{\gamma _0 }{\gamma }} \right)^{1 / 4}\cos \left( 
{\int_{ - R}^t {\Omega dt} } \right),
\label{trajectory}
\end{equation}
\noindent
Here, the energy of the electron is assumed to be $\gamma _0 $ at the time it enters the plasma cavity $\xi = - R$, $r_0 $ is the initial electron deviation from the cavity axis, and $\gamma _l = (1 - v_l^2 )^{ - 1 / 2}$ is the relativistic gamma factor of the laser pulse. Because of the electron acceleration, the betatron frequency $\Omega = 1 / \sqrt {2\gamma } $ is a slowly varying function of the time. The dependence of the electron energy on $\xi $ is defined by
\begin{equation}
\gamma \approx \gamma _0 + \frac{\gamma _l^2 }{2}\left( {R^2 - \xi ^2} 
\right).
\label{temp}
\end{equation}
\noindent
The electron energy increment,  $\gamma \approx \gamma _0 + \gamma _l^2 R^2 / 2$, , is at a maximum at the cavity center  ($\xi = 0)$. It follows from Eq.~(\ref{trajectory}) that the amplitude of the betatron electron oscillations decreases as the cavity center is approached. 

Making betatron oscillations, an ultrarelativistic electron emits electromagnetic waves and undergoes
recoil as a result of the photon emission. The relativistic equations of motion for an electron in an electromagnetic field with the radiative reaction force are \cite{landau}
\begin{eqnarray}
\frac{du^i}{ds} = F^{ik}u_k + \mu g^i, 
\label{reaction}
\\
g^i = \frac{\partial F^{ik}}{\partial x^l}u_k u^l - F^{il}F_{kl} u^k + 
\left( {F_{kl} u^l} \right)\left( {F^{km}u_m } \right)u^i,
\end{eqnarray}
\noindent
where $F_{ik} $ Fik is the electromagnetic field tensor, $u_k $ is the 4-velocity of the electron, $\mu = 2r_e \omega _p / (3c) $, $r_e = 3 \cdot 10^{13}$ρμ is the classical electron radius. The first term in
Eq.~(\ref{reaction}), corresponds to the Lorentz force and the second
term describes the action of the radiative reaction force.

Under our assumptions, the focusing forces (the transverse component of the Lorentz force) make a
major contribution to the energy losses through radiation, while the acceleration is attributable to the action of the longitudinal component of the Lorentz force. In this case, the approximate equations that describe the electron dynamics in the cavity are
\begin{eqnarray}
\frac{dp}{dt} = - \frac{y}{2} - \frac{\mu }{4}y^2p\gamma , \\
\label{approx1}
\frac{dy}{dt} = \frac{p}{\gamma }, \\
\label{approx2}
\frac{d\gamma }{dt} = - \frac{\xi }{2} - \frac{\mu }{4}y^2\gamma ^2.
\label{approx3}
\end{eqnarray}
\noindent
The first two equations describe the betatron electron oscillations. When the force of radiative friction is disregarded ($\mu = 0$) and for a slowly varying betatron frequency, the solution of the first two equations is (\ref{trajectory}). The first term on the right-hand side of the last equation describes the action of the longitudinal component of the Lorentz force that provides electron acceleration, while the second term describes the radiative losses.

Clearly, acceleration is possible as long as the accelerating force in the cavity is larger than the force of
radiative friction. The limiting energy of the electron at which its acceleration is possible can be easily estimated from Eq.~(\ref{approx3})
\begin{equation}
\label{limit1}
\frac{r_e }{3}\gamma ^2r_0^2 \frac{\omega _p^2 }{c^2} < R,
\end{equation}
\noindent
Inequality (\ref{limit1}) is written in dimensional units to take into account the influence of the plasma density in explicit form. It follows from the derived inequality that the square of the limiting energy of the electron at which its acceleration is still possible is proportional to the cavity size and inversely proportional to the plasma density and the square of the distance to the cavity axis. The stronger the focusing of the beam, the higher the energies to which it can be accelerated in the cavity.

Below, we will consider the two limiting cases. In the first case, the energy of the electron is relatively low and it makes many betatron oscillations while flying through the cavity. In the second case, at fairly high electron energy, the time of flight of the electron through the cavity is shorter than the betatron oscillation period.

\section*{3. ELECTRON DYNAMICS AT A LARGE NUMBER OF BETATRON OSCILLATIONS}

Let us first consider the case where the electron makes many betatron oscillations while flying through
one half of the cavity with an accelerating electric field. This condition limits the electron energy above,
\begin{equation}
\label{betatron-number}
\gamma < \frac{R^2}{2 \pi ^2}\gamma _l^4 .
\end{equation}
\noindent
For typical parameters of the interaction between an intense laser pulse and plasma \cite{Pukhov2002} $R \approx 5$, $\gamma _l \approx 10$ condition~(\ref{betatron-number}) is satisfied for electrons with energies
below $10$~GeV.

   \begin{figure}
   \includegraphics[width=8cm,clip]{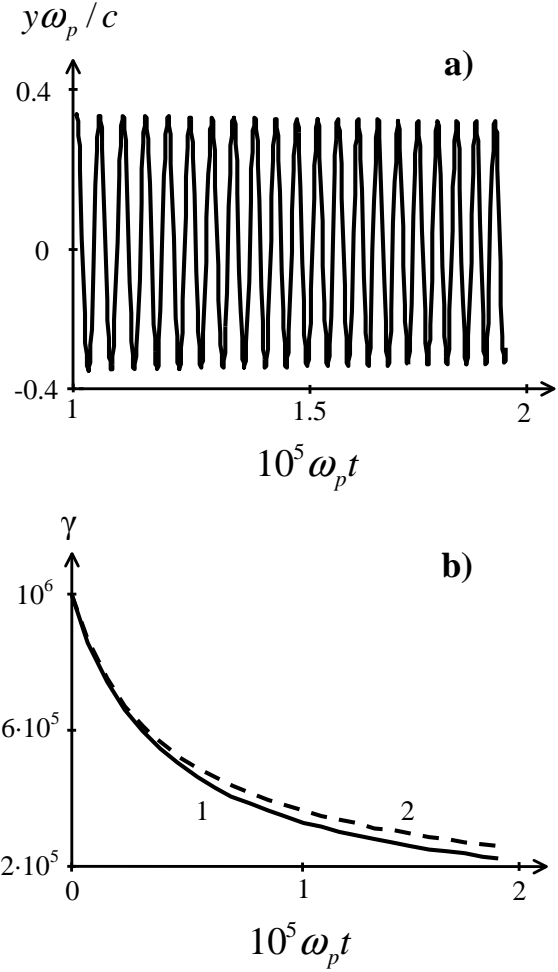}
   \caption
   {
   \label{Fig1}
   Transverse coordinate of the electron $y$ (a) and its energy $\gamma $, (b) versus time for the motion in an ion channel with the parameters $\gamma _0 = 5 \cdot 10^4$, $r_0 = 0.2$ θ $n_0 = 10^{19}$ρμ$^{- 3}$. Line~1 (solid) and line~2 (dashed) correspond to the approximate solution~(\ref{g-ion}) and the numerical solution of the equation
of motion~(\ref{reaction}), respectively.
}
   \end{figure}

When the number of betatron oscillations is large, it is convenient to use the averaging method~\cite{Bogolyubov}. To this end, let us introduce a new variable,
\begin{equation}
b\exp \left( {i\int \Omega dt} \right) = \frac{y}{2} - i\Omega p.
\label{b}
\end{equation}
\noindent
Substituting Eq.~(\ref{b}) in Eqs.~(\ref{approx1})-(\ref{approx3}) and averaging
over the phase yields the following reduced equations:
\begin{eqnarray}
\frac{db}{dt} = - \frac{\mu }{16}\gamma \left| b \right|^2b , 
\label{b-g1} \\
\frac{d\gamma }{dt} = \frac{1}{2}\left( {R - \frac{t}{2\gamma _l^2 }} 
\right) - \frac{\mu }{2}\left| b \right|^2\gamma ^2.
\label{b-g2}
\end{eqnarray}
\noindent
It should be noted that although rapidly oscillating component of $\gamma $ is small the derivative of it with respect to time can be large. So to derive Eq.~(\ref{b-g1}) we use Eq.~(\ref{approx3}) before averaging  in order to exclude $d \gamma / dt$. 

Assuming that $s = \left| b \right|^2 $, we can rewrite Eqs.~(\ref{b-g1}) and (\ref{b-g2}) as
\begin{eqnarray}
\frac{ds}{dt} = - \frac{\mu }{8 }\gamma s^2,
\label{bs1} \\
\frac{d\gamma }{dt} = \frac{1}{2}\left( {R - \frac{t}{2\gamma _l^2 }} 
\right) - \frac{\mu }{4}s\gamma ^2.
\label{bs2}
\end{eqnarray}
\noindent
where $r_0 = \sqrt {2\left\langle {y^2} \right\rangle } = 2\sqrt {s} $ is the amplitude of the betatron electron oscillations in the cavity. In the absence of an accelerating force, the equations take the form
\begin{eqnarray}
\frac{ds}{dt} = - \frac{\mu }{8}\gamma s^2,
\label{b-s1} \\
\frac{d\gamma }{dt} = - \frac{\mu }{4 }s \gamma ^2.
\label{b-s2}
\end{eqnarray}
\noindent
These equations describe the electron dynamics in an infinitely long ion channel where the longitudinal component
of the Lorentz force may be disregarded. The system of equations (\ref{b-g1}) and (\ref{b-g2}) has the integral of motion
\begin{equation}
\frac{|b|^4} {\sqrt{\gamma }}  = const = 
\frac{r_0^4 }{16 \sqrt{\gamma }},
\label{integral}
\end{equation}
\noindent
where the integration constant can be found from the initial conditions. It follows from Eq.~(\ref{integral}) that the betatron oscillation amplitude slowly decreasing as $y \propto \gamma ^{1/8}$ in this approximation. Thus, the radiated energy is pumped out from the longitudinal component as well as from transverse component of the electron energy. Solving Eqs. (\ref{b-g1}) and (\ref{b-g2}) yields 
\begin{equation}
\gamma = \frac{\gamma _0 }{(1 + 5 \mu r_0^2 t\gamma _0 / 32)^{4/5}}.
\label{g-ion}
\end{equation}
\noindent

As an example, let us consider the radiative deceleration of an electron with the initial parameters $\gamma _0 = 5 \cdot 10^4$, $r_0 = 0.2$ in an ion channel with the density $n_0 = 10^{19}$ρμ$^{- 3}$. As follows from Fig.~1, the constructed solution is in good agreement with the results of our numerical
integration of the equation of motion (\ref{reaction}).

Let us now consider the electron dynamics in a plasma cavity by taking into account the accelerating force and the radiative reaction force. Let us assume,that the betatron oscillation amplitude is approximately constant with time. This assumption is valid when the energy increment of the electron as a result of its acceleration is considerably smaller than its initial energy ($\Delta \gamma / \gamma \approx \gamma _l^2 R^2 / \left( {4\gamma } \right) \ll 1$). Integrating Eq.~(\ref{bs2}) with the condition $\left\langle {y^2} \right\rangle = r_0^2 / 2 = const$ yields
\begin{equation}
\gamma = \frac{4\alpha }{\mu r_0^2 \gamma _l^2 }\frac{Ai'\left( {\alpha \xi 
} \right) + \delta Bi'\left( {\alpha \xi } \right)}{Ai\left( {\alpha \xi } \right) 
+ \delta Bi\left( {\alpha \xi } \right)},
\label{accel1}
\end{equation}
\noindent
where $\xi = - R + t / (2\gamma _l^2 )$, $Ai(x)$ and $Bi(x)$ are the Airy functions \cite{Abramovitz} , $Ai'(x)$ and $Bi'(x)$ are the derivatives of the Airy functions, and $\alpha = - \gamma _l^{4 / 3} r_0^{2 / 3} \mu ^{1 / 3}2^{ - 2 / 3}$, $\delta$ is the integration constant, which can be found from the initial condition $\gamma \left( {\xi = - R} \right) = \gamma _0 $. It should be noted that the factor $\alpha $ does not depend on the electron energy. Moreover, for typical laser-plasma interaction parameters,  $\alpha \xi \ll 1$, and the asymptotics of the Airy
function may be used for low values of the argument. Expanding separately the numerator and the denominator in Eq.~(\ref{accel1}) in terms of the small parameter $\alpha \xi$, we obtain the following expression for the electron energy:
\begin{equation}
\gamma \approx \frac{\gamma _0 \left( {32 + \mu Rt^2r_0^2 } \right) + 
4\gamma _l^{ - 2} t\left( {4\gamma _l^2 R - t} \right)}{32 + \mu tr_0^2 
\left( {4\gamma _0 + tR} \right)}.
\label{accel2}
\end{equation}

Let us consider the electron dynamics in a plasma cavity with the parameters $\gamma _0 = 4 \cdot 10^4$, $\gamma _l = 10$, $R = 15$, $r_0 = R / 4$, $n_0 = 10^{19}$ρμ$^{ - 3}$. Figure~2 presents the time dependences of the electron density obtained by integrating the equation of motion (\ref{reaction}) with and without the radiative reaction force. At $t = 3000 $, the electron reaches the cavity center, where the accelerating force vanishes. As we see from Fig.~2, when the radiative losses are disregarded, the electron energy reaches its maximum at the cavity center. Including the radiative
losses leads to a significant reduction in the maximum electron energy. It follows from Fig.~2 that the approximate
solution(\ref{accel2}) is in good agreement with the results of our numerical integration of the equation of motion.

   \begin{figure}
   \includegraphics[width=8cm,clip]{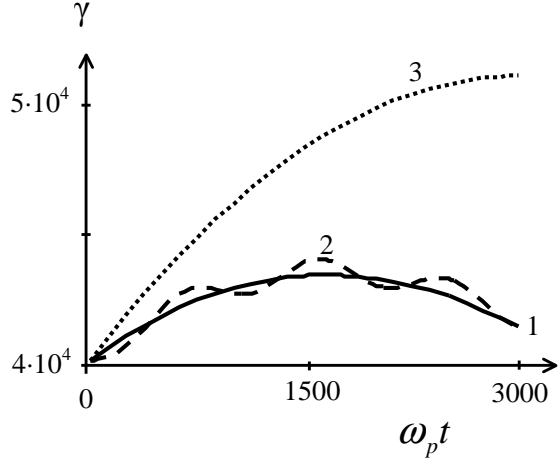}
   \caption
   {
   \label{Fig2}
   Electron energy versus time for the motion in a plasma channel with the parameters $\gamma _0 = 4 \cdot 10^4$, 
$\gamma _l = 10$, $R = 15 c/\omega _p$, $r_0 = R / 4$, $n_0 =  10^{19}$ρμ$^{ - 3}$.  The solid, dashed, and dotted lines correspond to the approximate solution~(\ref{accel2}), the numerical solution of the equation of motion~(\ref{reaction}),
and the numerical solution of the equation of motion~(\ref{reaction})
without the action of radiative friction ($\mu = 0$), respectively.
}
   \end{figure}

\section*{4. ELECTRON DYNAMICS IN THE ABSENCE OF BETATRON OSCILLATIONS}

In the opposite limit of high electron energies ($\gamma > R^2\gamma _l^4 / \pi ^2$) without the action of the radiative force, the electron makes no more than one betatron oscillation while flying through the accelerating half of the cavity.
Let us assume that the accelerating force in the cavity is larger than or comparable to the radiative reaction force
in the longitudinal direction, $\gamma ^2 < 2R / \left( {\mu r_0^2 } \right)$ . Since the energy increment during the acceleration at high electron energies is relatively small, the electron energy in the equations for the transverse coordinate (\ref{approx1}) and (\ref{approx2}) may be considered as a constant. Thus, for the transverse coordinate of the electron, we derive the expression
\begin{equation}
y\left( t \right) \approx r_0 \cos \Omega t.
\label{y1}
\end{equation}
\noindent
Substituting this expression in Eq.~(\ref{approx3}) yields an equation for the electron energy:
\begin{equation}
\frac{d\gamma }{dt} = \frac{1}{2}\left( {R - \frac{t}{2\gamma _l^2 }} 
\right) - \frac{\mu }{4}\gamma ^2r_0^2 \cos ^2\Omega t.
\label{g1}
\end{equation}
\noindent
For the case where $\Omega t \ll 1$ , the solution of this equation was found in the previous section and is represented by Eqs.~(\ref{accel1}) and (\ref{accel2}), in which the substitution $r_0^2 \to 2r_0^2 $ should be made.

As the initial electron energy increases, the accelerating force can become negligible compared to the force of radiative friction ($\gamma ^2 \gg 2R\mu ^{ - 1}r_0^{ - 2} )$. In this case, the energy losses through radiation can be significant. The betatron oscillation amplitude increases as the electron energy decreases as a result of the generation
of electromagnetic radiation and the time dependence of the transverse coordinate should be taken into account. By disregarding the accelerating force, we can reduce Eqs.~(\ref{approx1})-(\ref{approx3}) to the form
\begin{eqnarray}
\frac{d^2y}{dt^2} + \frac{y}{2\gamma } = 0,
\label{y2} \\
\frac{d\gamma }{dt} = - \frac{\mu }{4}y^2\gamma ^2
\label{g2}.
\end{eqnarray}
It should be noted that in the absence of an accelerating force, the terms in Eq.~(\ref{y2}) that describe the action of
the nonlinear force of friction cancel each other out.

In the zeroth approximation, we will assume that the transverse coordinate does not depend on the time. The time evolution of the electron energy in the cavity is then defined by Eq.~(\ref{g-ion}). In the next approximation, we
will take into account the time dependence of the electron energy in Eq.~(\ref{y2}). As a result, we obtain the following
equation for the transverse coordinate:
\begin{equation}
\frac{d^2y}{dt^2} + \left( {\frac{1}{2\gamma _0 } + \frac{\mu r_0^2 }{8}t} 
\right)y = 0 ,
\label{y3}
\end{equation}
\noindent
Its solution can be expressed in terms of the Airy functions: 
\begin{eqnarray}
y = \pi r_0 \left[ {Ai\left( \tau \right)Bi'\left( \varepsilon \right) - 
Bi\left( \tau \right)Ai'\left( \varepsilon \right)} \right],
\label{y4} \\
\tau = \varepsilon \left( {1 + \beta t} \right),
\label{tau}
\end{eqnarray}
\noindent
where $\varepsilon = - 2\mu ^{ - 2 / 3}\gamma _0^{ - 1} r_0^{ - 4 / 3} $, $\beta = \mu \gamma _0 r_0^2 / 4$.
Substituting the solution obtained in Eq.~(\ref{g2}) yields an equation for the electron energy. The solution of this
equation can also be expressed in terms of the Airy functions:

\begin{equation}
\frac{1}{\gamma } - \frac{1}{\gamma _0 } = \frac{\pi ^2}{\gamma _0 
\varepsilon }\left\{ {\tau \left[ {Ai\left( \tau \right)Bi'\left( 
\varepsilon \right) - Bi\left( \tau \right)Ai'\left( \varepsilon \right)} 
\right]^2 + \left[ {Ai'\left( \tau \right)Bi'\left( \varepsilon \right) - 
Bi'\left( \tau \right)Ai'\left( \varepsilon \right)} \right]^2 - 
\frac{\varepsilon }{\pi ^2}} \right\}.
\label{g3}
\end{equation}
\noindent
In the limit $\beta t \ll 1$, the solution obtained transforms to Eq.~(\ref{g-ion}).

As an example, let us consider the motion of an electron with the initial energy $\gamma _0 = 4 \cdot 10^4$ and the initial deviation from the axis $r_0 = 0.1$. The ambient plasma density is $n_0 = 10^{19}$ρμ$^{ - 3}$.  Figure~3 shows the dynamics of the transverse coordinate of the electron and its energy for the motion in a cavity with the reaction force. As follows from Fig.~3, the derived expressions~(\ref{y4}) and (\ref{g3}) are in good agreement with the results of our numerical integration of the equations of motion.

   \begin{figure}
   \includegraphics[width=8cm,clip]{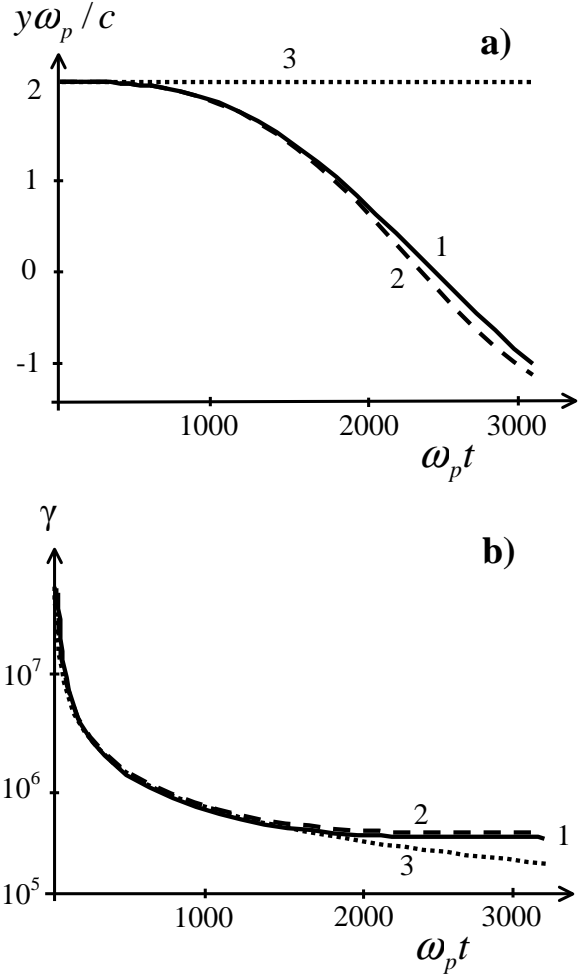}
   \caption
   {
   \label{Fig3}
   Transverse coordinate of the electron (a) and its energy (b) versus time for the motion in a plasma cavity with
the parameters $\gamma _0 = 5 \cdot 10^4$, $r_0 = 0.2$ θ $n_0 = 10^{19}$ρμ$^{- 3}$. The solid, dashed, and dotted lines correspond to the approximate solution~(\ref{y4}) and (\ref{g3}), the numerical solution of the equation of motion~(\ref{reaction}), and the approximate solution~(\ref{g-ion})
and $y = $const, respectively.
}
   \end{figure}

\section*{5. RESULTS OF NUMERICAL SIMULATIONS}

The action of radiative friction on the external beam of relativistic electrons accelerated in a plasma cavity was numerically simulated using a two-dimensional relativistic hybrid particle-in-cell code for a cylindrical geometry \cite{kostyukov-kod}. The quasi-static approximation (the plasma response is assumed to vary slowly during the characteristic evolution time of the laser pulse) is used to speed up the computation. The code includes the action of the force of radiative friction on the electron dynamics. 

In our numerical simulations, we assumed that the laser pulse was circularly polarized and had a Gaussian profile,
$a =  a_0 \exp \left( { - r^2 / r_l^2 - \xi ^2 / L_l ^2} \right)$, and the wavelength $\lambda = 0.82$~µm. The laser pulse parameters are $r_l = 5$, $L_l = 2$, $a_0 = 10$. The pulse propagates in a plasma with the density $n_0 = 10^{-19}$ cm$^{ - 3}$. The beam parameters are the following: the relativistic gamma factor of the electrons is $\gamma = 10^5$, the beam radius is $r_b = 2$, and the beam electron density is $n_b = 10^{ - 17}$ cm$^{ - 3}$.

Figure~4a shows the distribution of the electron plasma and beam density. The beam electrons are accelerated in a plasma cavity with the radiative reaction force. The transverse forces in the cavity focus the beam. As follows from Fig.~4a, the beam electrons reach the cavity center approximately in half the betatron period, which matches the estimate of the time it takes for the electron to reach the cavity center. 

The beam electron distribution function at the time the leading edge of the beam reaches the cavity center is shown in Fig.~4b (line~1). At the beginning of the interaction, all of the beam electrons had the same energy (line~2 in Fig.~4b). As follows from Fig.~4b, the radiative reaction force decelerates the beam electrons with a large betatron oscillation amplitude, while the electrons with a small amplitude are accelerated by the longitudinal electric field. According to our numerical simulations, the energy of the electrons with a small betatron oscillation amplitude increases by $1$~GeV, which is close to the estimate that follows from Eq.~(\ref{temp}). The force of radiative friction for the electrons with the maximum oscillation amplitude ($r_0 \simeq 2$) exceeds the accelerating force; as a result, these electrons are decelerated, losing $2$~GeV of their energy (see Fig.~4b). A similar estimate ($2.5$~GeV) for the energy lost by the electron through radiation is obtained from Eq.~(\ref{accel2}),
where it is considered that $t \simeq \pi \sqrt{2 \gamma}$.

   \begin{figure}
   \includegraphics[width=8cm,clip]{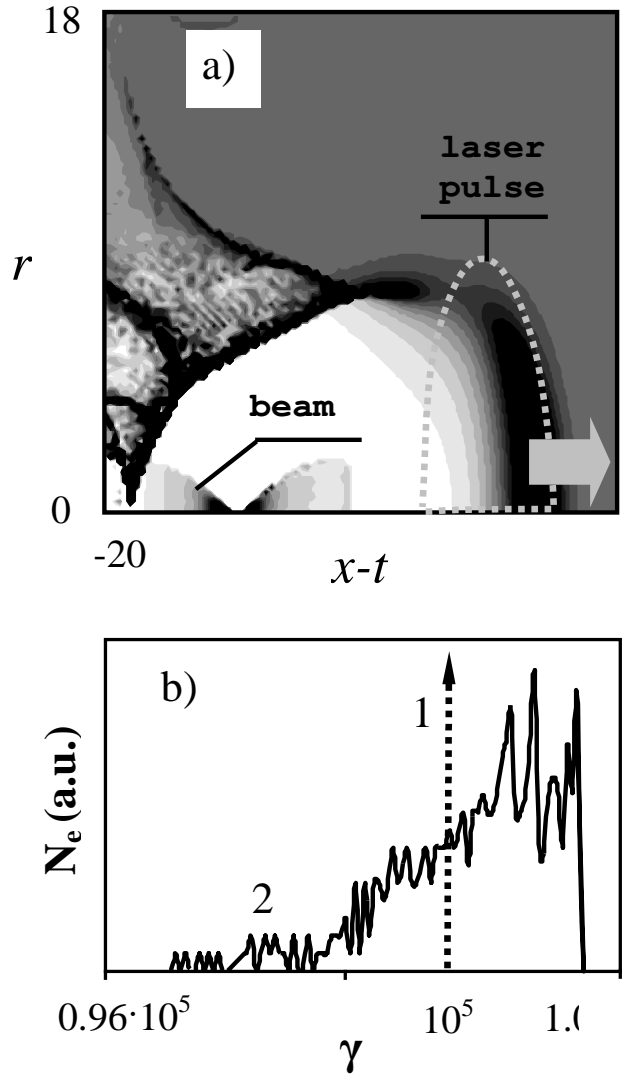}
   \caption
   {
   \label{Fig4}
   (a) Plasma electron and electron beam density. Darker colors correspond to higher electron densities. The location of the laser pulse is indicated by the dotted line. The coordinates are presented in units of $c/\omega _{p}$. (b) The beam electron energy distribution function: at the initial time (dashed line 1) and at the time the leading edge of the beam reached the cavity center (solid line 2).
}
   \end{figure}

\section*{6. CONCLUSIONS}

Figure~5 shows the various regimes of electron dynamics in a plasma cavity with the radiative reaction
force in the $\gamma _0$ (initial electron energy)- $r_0$ (initial deviation) plane. As follows from Fig.~5 and inequality (\ref{limit1}), the force of radiative friction can restrict significantly the electron
acceleration in plasma. The radiative friction can be reduced by reducing the beam radius and the plasma
density. For the plasma density $n_0 = 10^{19}$ρμ$^{ - 3}$, electron acceleration is not possible at energies above $50$~GeV and at a radius of the accelerated bunch $r_0 > 10$~µm, which is typical of present-day accelerators \cite{clark}. However, the projected accelerators (SLAC, TESLA) are expected to provide beam focusing up to tens or hundreds of nanometers \cite{SLAC,TESLA}. In this case, electron acceleration is possible up to several tens of TeV.

   \begin{figure}
   \includegraphics[width=8cm,clip]{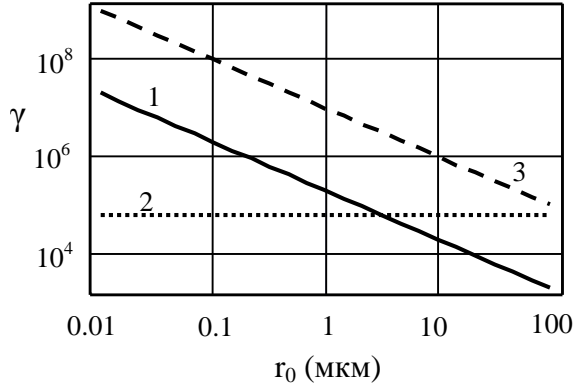}
   \caption
    {
    \label{Fig5}
    Various regimes of electron dynamics in a plasma cavity with the radiative reaction force in the $\gamma _0$ (initial electron energy)-$r_0$ (initial deviation) plane. Straight line~1 corresponds to the boundary at which the accelerating force is equal to the force of radiative friction. Straight line~2 corresponds to the boundary at which the time of flight through the cavity is equal to the betatron oscillation period. Straight line~3 corresponds to the boundary that separates the quantum domain of radiation from the classical one. The interaction parameters are $\gamma _l = 10$, $R = 8 c/\omega _p$, $n_0 = 10^{19}$ρμ$^{ - 3}$.}
   \end{figure}

Since the radiated photon energy increases quadratically with electron energy, these energies can become
equal for ultrarelativistic electrons. In this case, the radiation and the electron dynamics itself are of a quantum nature. As follows from Fig.~4, for $n_0 = 10^{19}$cm$^{ - 3}$ and $r_0 > 10$, the quantum effects become significant for electrons with energies above $500$~GeV. Thus, the quantum theory should be used at high electron energies. In recent years, impressing results on the acceleration of electrons in plasma by relativistic electron bunches have been obtained \cite{gev}. The acceleration is most efficient in the ion-focusing regime where the bunch density is higher than the ambient plasma density \cite{blowout}. In this case, the plasma electrons are expelled almost completely from the region around the bunch by  its large negative charge. The electromagnetic field distribution in this region is identical to the field distribution
inside the plasma cavity generated by a short laser pulse \cite{Mori}. Thus, the results obtained are also valid for describing the dynamics of the electrons accelerated by a relativistic electron beam in plasma.

\end{document}